\documentclass[lettersize,journal]{IEEEtran}
\usepackage{amsmath,amsfonts}
\usepackage{algorithmic}
\usepackage{algorithm}
\usepackage{array}
\usepackage[caption=false,font=normalsize,labelfont=sf,textfont=sf]{subfig}
\usepackage{textcomp}
\usepackage{stfloats}
\usepackage{url}
\usepackage{verbatim}
\usepackage{graphicx}
\usepackage{cite}
\usepackage{steinmetz}
\usepackage{xcolor}
\usepackage{pgfplots}
\usepackage{booktabs}
\allowdisplaybreaks
\usepackage{enumitem}
\usepackage{algorithm}
\usepackage{algorithmic}

\begin{document}

\title{Beyond the Neural Fog: Interpretable Learning for AC Optimal Power Flow}

\author{S. Pineda,~\IEEEmembership{Senior Member,~IEEE,} J. Pérez-Ruiz,~\IEEEmembership{Member,~IEEE,}, J. M. Morales,~\IEEEmembership{Senior Member,~IEEE,} 
\thanks{The work of S. Pineda and J. M. Morales was supported in part by the Spanish Ministry of Science and Innovation (AEI/10.13039/501100011033) through projects PID2020-115460GB-I00 and PID2023-148291NB-I00. S. Pineda, J. Pérez-Ruiz and J. M. Morales are with the research group OASYS, University of Malaga, Malaga 29071, Spain: spineda@uma.es; jperez@uma.es; juan.morales@uma.es. Finally, the authors thankfully acknowledge the computer resources, technical expertise, and assistance provided by the SCBI (Supercomputing and Bioinformatics) center of the University of M\'alaga.}
}



\maketitle

\begin{abstract}
The AC optimal power flow (AC-OPF) problem is essential for power system operations, but its non-convex nature makes it challenging to solve. A widely used simplification is the linearized DC optimal power flow (DC-OPF) problem, which can be solved to global optimality, but whose optimal solution is always infeasible in the original AC-OPF problem. Recently, neural networks (NN) have been introduced for solving the AC-OPF problem at significantly faster computation times. However, these methods necessitate extensive datasets, are difficult to train, and are often viewed as black boxes, leading to resistance from operators who prefer more transparent and interpretable solutions. In this paper, we introduce a novel learning-based approach that merges simplicity and interpretability, providing a bridge between traditional approximation methods and black-box learning techniques. Our approach not only provides transparency for operators but also achieves competitive accuracy. Numerical results across various power networks demonstrate that our method provides accuracy comparable to, and often surpassing, that of neural networks, particularly when training datasets are limited.
\end{abstract}


\section{Introduction}\label{sec:introduction}
\IEEEPARstart{T}{he} Optimal Power Flow (OPF) is a fundamental cornerstone for ensuring the secure, reliable, and efficient operation of power systems. It serves as a critical decision-making tool for both power system planning and real-time operations. The primary objective of the OPF is to determine the optimal output of power plants to meet electricity demand at the minimum cost while satisfying the technical constraints of the network. However, solving the OPF problem presents significant challenges due to its non-convex nature. This non-convexity complicates the search for a globally optimal solution, as optimization algorithms can easily become trapped in local minima. Besides, traditional optimization algorithms applied to AC-OPF problems often struggle with convergence and may have difficulties finding an optimal solution, whether local or global, within a reasonable number of iterations. This concern is especially critical for practical applications such as day-ahead market clearing, where reliable convergence criteria are essential for determining electricity prices. Moreover, solving AC-OPF involves large-scale nonlinear optimization, which is computationally intensive, especially for large real-world networks. Efficiently handling these large-scale instances often requires advanced optimization techniques and substantial computing resources. The lack of convergence guarantees and the high computational burden represent major barriers to using AC-OPF models in real-time operations, where rapid and accurate solutions are crucial due to the dynamic nature of power system conditions \cite{nair2022computational}. 

The traditional approach to solving the OPF problem involves a DC approximation, which is formulated as a linear programming problem with a global optimal solution \cite{frank2016introduction}. This method has been widely adopted and underpins many market-clearing algorithms. However, its limitations in capturing the complexities of power systems are becoming increasingly apparent. While the DC approximation offers computational efficiency, it often leads to significant economic losses and suboptimal solutions. More importantly, the solution provided by the DC approximation is always AC-infeasible \cite{baker2021solutions}. The growing prevalence of reactive power and voltage constraints in contemporary networks further exacerbates the accuracy concerns associated with this approach \cite{yang2017solving}. The technical literature also proposes other linearized versions of the OPF problem beyond the traditional DC-OPF. For instance, \cite{yang2017linearized} introduces a linearization of OPF based on loss factors. Similarly, \cite{misra2018optimal} presents a bilevel problem to find the best parameters to approximate OPF with linear equations.


Alternatively, relaxed formulations using second-order cone or semidefinite programming have been developed \cite{low2014convex}. The advantage of these approaches lies in their convexity, which guarantees that they can be solved to global optimality. However, these methods are still computationally demanding and since they have a larger feasible set, they may deliver solutions that are AC-infeasible. Besides, it has been documented that the optimality gaps for these methodologies can be significant for certain networks \cite{yang2017solving}. 

More recently, machine learning techniques have been introduced to solve the OPF problem \cite{khaloie2024review}. Deep neural networks have been employed to directly learn the solution of the DC-OPF \cite{pan2020deepopf} or to learn the active set of constraints in AC-OPF \cite{zamzam2020learning}, while sensitivity-informed, physics-informed and graph neural networks have been proposed to address the AC-OPF problem \cite{singh2021learning, owerko2020optimal, lopez2024optimal, lei2020data}. These methods offer rapid computation, facilitating real-time implementation. However, they typically require extensive datasets and are challenging to train. Moreover, the black-box nature of neural networks raises concerns about transparency and interpretability, making system operators hesitant to deploy them in real networks.


In this paper, we propose a learning-based methodology to reformulate the OPF problem into a convex optimization problem that closely approximates the AC-OPF solution. Our approach combines model- and learning-based methodologies by utilizing a first-order Taylor expansion of the original AC power flow equations around an estimated operating point derived from historical data. Similar to other model-based approximations, a global optimal solution to the convex programming model we propose can be achieved using standard optimization algorithms. However, unlike the well-known DC-OPF approximation, our approach adapts to each operating point, resulting in solutions that are closer to those of the original AC-OPF problem. Similar to other learning-based approaches, ours leverages historical data to improve accuracy. However, unlike approaches based on neural networks, our method is easy to train, transparent, and interpretable, making it more suitable for adoption in real-world power systems.

\begin{figure}
    \centering
    \includegraphics[width=0.64\linewidth]{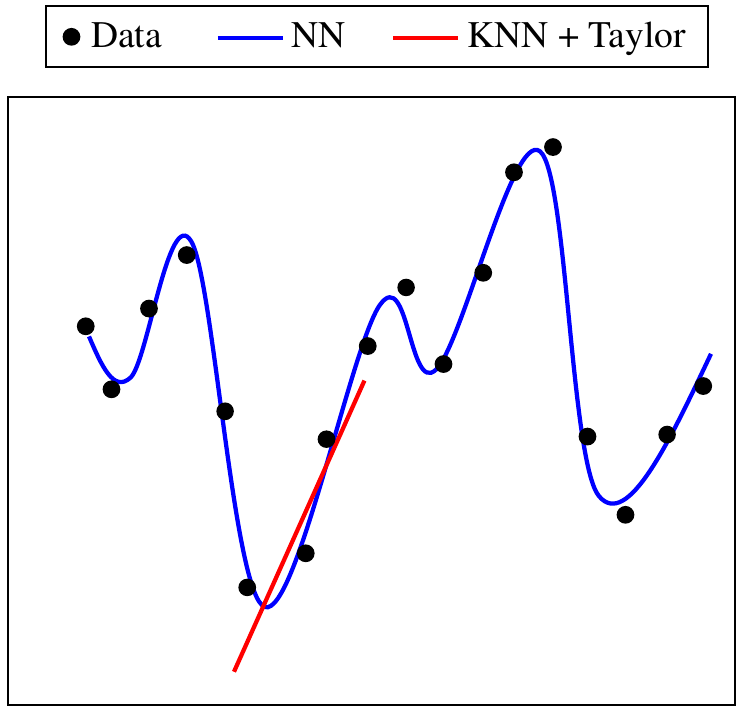}
        \caption{Visual illustration of how neural networks and our approach work}
    \label{fig:nn_vs_proposed}
\end{figure}

Figure \ref{fig:nn_vs_proposed} visually compares the \emph{modus operandi} of  neural networks and our approach. Neural networks (NN) seek to learn a complex nonlinear function that fits all observed data points, represented as black dots. In contrast, our method (KNN+Taylor) focuses on a localized linear approximation: By employing a $K$-nearest neighbor (KNN) clustering algorithm, we identify the region of interest centered around the anticipated optimal power flow solution and, within this region, we apply a Taylor expansion to linearly approximate the power flow equations. Since the neural network aims to approximate the function across its entire domain, it requires a large amount of training data and is challenging to train. In contrast, the proposed approach requires significantly less data by approximating the function locally and dynamically adapting it to each specific instance.


    
         
    

The main contributions of this paper are threefold. First, we propose a hybrid approach to infer the solution of the AC-OPF problem by combining model- and learning-based methodologies. This approach uses a first-order Taylor expansion of the non-convex power flow equations around an operating point inferred from historical data.  We present two variants of our approach: the first focuses solely on the relationship between active power flows and voltage angles, similar to the traditional DC-OPF, while the second includes active and reactive power as well as voltage angles and magnitudes, aligning more closely with the original AC-OPF. Second, we evaluate the performance of our methodology against the classical model-based DC-OPF approximation and two learning-based approaches: a simple and fast non-parametric nearest neighbor regression, and a more complex deep neural network regression. The evaluation metrics include a statistical measure based on mean squared error, an infeasibility metric of the obtained solution, and a metric assessing the redispatch actions required to achieve feasibility. Finally, we provide numerical results comparing these methodologies across four different power systems commonly used in the literature, utilizing a benchmark database for the AC-OPF problem. 

The rest of the paper is organized as follows. Section \ref{sec:formulations} introduces the notation and formulates the various optimization models used in this study. Section \ref{sec:methodology} details the different learning-based methodologies employed to infer the solution of the AC-OPF problem. The comparison metrics used to evaluate the performance of these approaches are outlined in Section \ref{sec:comparison}. Section \ref{sec:simulations} presents and analyzes the numerical results. Finally, Section \ref{sec:conclusions} draws the conclusions of this work.

\section{Formulations}\label{sec:formulations}

Consider a power network composed of nodes, branches connecting the nodes, and generators. The nodes, indexed by $n \in \mathcal{N}$, are characterized by a shunt conductance ($G^{sh}_n$), a shunt susceptance ($B^{sh}_n$), an active power demand ($p^d_n$), and a reactive power demand ($q^d_n$). Given the inherent complexity of the optimal power flow in its deterministic formulation, we assume throughout this work that electricity demand is known with certainty. Addressing the stochastic version of this problem is considered beyond the scope of this study. The voltage magnitude and angle at each node are denoted by $v_n$ and $\theta_n$, respectively, and the fomer is bounded by technical limits $\underline{v}_n$, $\overline{v}_n$. The generators, indexed by $g \in \mathcal{G}$, are characterized by a convex cost function $c_g()$ and minimum/maximum active and reactive power outputs denoted by $\underline{p}_g/\overline{p}_g$ and $\underline{q}_g/\overline{q}_g$, respectively. The branches, indexed by $l \in \mathcal{L}$, are characterized by a series resistance ($r_l$), a series reactance ($x_l$), a total charging susceptance ($b_l$), a tap ratio magnitude ($\tau_l$), and a phase shift angle ($\gamma_l$). We assume all branch parameters are known. We define two sparse connection matrices $F$ and $T$ of size $|\mathcal{L}| \times |\mathcal{N}|$ such that $F_{ln}$ and $T_{lm}$ are equal to 1 if node $n/m$ is the start/end node of branch $l$, with all other elements being zero. The active power flow from/to each branch $l$ are denoted by $p^f_l$ and $p^t_l$, respectively, while $q^f_l$ and $q^t_l$ similarly represent the reactive power flows. For each branch the apparent power flow is limited by $\overline{s}_l$ and we also define:
\begin{subequations}
\begin{align}
& G^{ff}_l + jB^{ff}_l = \left(\tfrac{1}{r_l+jx_l}+j\tfrac{b_l}{2} \right)\tfrac{1}{\tau_l^2} \\
& G^{ft}_l + jB^{ft}_l = -\tfrac{1}{r_l+jx_l}\tfrac{1}{\tau_l e^{-j\gamma_l}} \\
& G^{tf}_l + jB^{tf}_l = -\tfrac{1}{r_l+jx_l}\tfrac{1}{\tau_l e^{j\gamma_l}} \\
& G^{tt}_l + jB^{tt}_l = \left(\tfrac{1}{r_l+jx_l}+j\tfrac{b_l}{2} \right)
\end{align}
\end{subequations}
Using this notation, the AC-OPF problem can be formulated as the following optimization model \cite{nair2022computational, bienstock2022mathematical}:
\begin{subequations}
\begin{align}
& \min_{p_g,p^f_l,q^f_l,p^t_l,q^t_l,v_n,\theta_n} \sum_g c_g(p_g)  \label{eq:ac_opf_of}\\
& \text{subject to} \nonumber\\
& \sum_{g\in\mathcal{G}_n} p_g - p^d_n = v_n^2G^{sh}_n + \sum_{l} (F_{ln}p^f_l + T_{ln}p^t_l), \forall n\label{eq:ac_opf_balp}\\
& \sum_{g\in\mathcal{G}_n} q_g - q^d_n = -v_n^2G^{sh}_n + \sum_{l} (F_{ln}q^f_l + T_{ln}q^t_l), \forall n\label{eq:ac_opf_balq}\\
& p^f_l = v^2_{n}G^{ff}_l + v_{n}v_{m}(G^{ft}_l\cos(\theta_{nm})+B^{ft}_l\sin(\theta_{nm})), \nonumber \\
& \qquad \forall(l,n,m):F_{ln}=T_{lm}=1\label{eq:ac_opf_pf}\\
& q^f_l = -v^2_{n}B^{ff}_l + v_{n}v_{m}(G^{ft}_l\sin(\theta_{nm})-B^{ft}_l\cos(\theta_{nm})), \nonumber \\
& \qquad \forall(l,n,m):F_{ln}=T_{lm}=1\label{eq:ac_opf_qf}\\
& p^t_l = v^2_{m}G^{tt}_l + v_{n}v_{m}(G^{tf}_l\cos(\theta_{mn})+B^{tf}_l\sin(\theta_{mn})), \nonumber \\
& \qquad \forall(l,n,m):F_{ln}=T_{lm}=1\label{eq:ac_opf_pt}\\
& q^t_l = -v^2_{m}B^{tt}_l + v_{n}v_{m}(G^{tf}_l\sin(\theta_{mn})-B^{tf}_l\cos(\theta_{mn})), \nonumber \\
& \qquad \forall(l,n,m):F_{ln}=T_{lm}=1\label{eq:ac_opf_qt}\\
& \underline{p}_g \leq p_g \leq \overline{p}_g, \forall g \label{eq:ac_opf_pglim}\\
& \underline{q}_g \leq q_g \leq \overline{q}_g, \forall g \label{eq:ac_opf_qglim}\\
& \underline{v}_n \leq v_n \leq \overline{v}_n, \forall n \label{eq:ac_opf_vlim}\\
& (p^f_l)^2 + (q^f_l)^2 \leq (\overline{s}_l)^2, \forall l \label{eq:ac_opf_sfmax}\\
& (p^t_l)^2 + (q^t_l)^2 \leq (\overline{s}_l)^2, \forall l \label{eq:ac_opf_stmax}
\end{align} \label{eq:ac_opf}
\end{subequations}
\noindent where $\mathcal{G}_n$ is the set of generators connected to node $n$, $\theta_{nm}=\theta_n - \theta_m$, and $\theta_{mn}=\theta_m - \theta_n$. The objective function \eqref{eq:ac_opf_of} aims to minimize the total operating cost of the system. Equations \eqref{eq:ac_opf_balp} and \eqref{eq:ac_opf_balq} enforce, respectively, the active and reactive power balances at each node $n$. The active and reactive power flows through each branch $l$ are calculated using the equality constraints \eqref{eq:ac_opf_pf}--\eqref{eq:ac_opf_qt}. Finally, technical limits are enforced by constraints \eqref{eq:ac_opf_pglim}--\eqref{eq:ac_opf_stmax}. Although objective function \eqref{eq:ac_opf_of} is usually quadratic, linear, or piece-wise linear, model \eqref{eq:ac_opf} is non-convex due to constraints \eqref{eq:ac_opf_pf}--\eqref{eq:ac_opf_qt}. 

The DC approximation simplifies the problem by assuming that voltage magnitudes are close to 1 p.u., branch reactances are significantly higher than branch resistances, and voltage angle differences are small \cite{stott2009dc}. Under these assumptions, the DC-OPF is formulated as follows:
\begin{subequations}
\begin{align}
& \min_{p_g,p^f_l,\theta_n} \sum_g c_g(p_g) \label{eq:dc_opf_of}\\
& \text{subject to} \nonumber\\
& \sum_{g\in\mathcal{G}_n} p_g - p^d_n = \sum_{l} (F_{ln}p^f_l - T_{ln}p^f_l), \forall n\\
& p^f_l = \tfrac{\theta_{n} - \theta_{m}}{x_l}, \forall(l,n,m):F_{ln}=T_{lm}=1 \label{eq:dc_opf_pf}\\
& \underline{p}_g \leq p_g \leq \overline{p}_g, \forall g \\
& -\overline{s}_l \leq p^f_l \leq \overline{s}_l, \forall l 
\end{align} \label{eq:dc_opf}
\end{subequations}
Model \eqref{eq:dc_opf} is convex and therefore can be solved to global optimality with off-the-shelf optimization software. As discussed in \cite{stott2009dc}, the DC formulation of the OPF offers several advantages. Firstly, its solutions are non-iterative, reliable, and unique, ensuring consistency and dependability in results. Secondly, the methods and software used in this formulation are relatively simple and user-friendly. Thirdly, the DC-OPF can be solved and optimized efficiently, which is particularly beneficial in the demanding area of contingency analysis. Additionally, it requires minimal network data that is relatively easy to obtain. The convex nature of the DC-OPF aligns well with the economic theories on which the majority of current transmission-constrained electricity markets are based. Lastly, the approximated MW flows provided by the DC-OPF are reasonably accurate, especially for the heavily loaded branches that might constrain system operation.

However, the assumption that the slope of this linear function should necessarily equal the branch susceptance is questionable, as discussed in \cite{shao2021data, taheri2024optimizing}, where the authors propose other data-driven methodologies to compute the intercept and slope that better approximate the linear relationship between active power flows and voltage angle differences. Moreover, as highlighted by \cite{baker2021solutions}, DC solutions are never AC-feasible, even with generation adjustments in DC-OPF to account for losses. Finally, as demonstrated in \cite{li2022numerical}, the accuracy of these model-based linear approximations of the OPF problem are very system and case dependent.

As proposed in \cite{coffrin2014linear,yang2018general}, equations \eqref{eq:ac_opf_pf}--\eqref{eq:ac_opf_qt} can be also linearized using first-order Taylor expansion around a given operating point $\{\tilde{v}_n, \tilde{\theta}_n, \forall n\}$, leading to the following formulation of the optimal power flow problem: 
\begin{subequations}
\begin{align}
& \min_{p_g,p^f_l,q^f_l,p^t_l,q^t_l,v_n,\theta_n} \sum_g c_g(p_g) \\
& \text{subject to} \nonumber\\
& \hspace{-2mm} \sum_{g\in\mathcal{G}_n} \hspace{-1mm} p_g \!-\! p^d_n \!=\! \tilde{v}_nG^{sh}_n(2v_n\!-\!\tilde{v}_n) \!+\!\hspace{-1mm} \sum_{l} (F_{ln}p^f_l \!+\! T_{ln}p^t_l), \forall n\\
& \hspace{-2mm} \sum_{g\in\mathcal{G}_n} \hspace{-1mm} q_g \!-\! q^d_n \!=\! \tilde{v}_nB^{sh}_n(\tilde{v}_n\!-\!2v_n)  \!+\!\hspace{-1mm} \sum_{l} (F_{ln}q^f_l \!+\! T_{ln}q^t_l), \forall n\\
& \hspace{-2mm} p^f_l \!-\! \tilde{p}^f_l \!=\! \tfrac{\partial p^f_l}{\partial v_n}(v_n\!-\!\tilde{v}_n) \!+\! \tfrac{\partial p^f_l}{\partial v_m}(v_m\!-\!\tilde{v}_m) \!+\! \tfrac{\partial p^f_l}{\partial \theta_{nm}}(\theta_{nm}\!-\!\tilde{\theta}_{nm}), \nonumber \\
&\qquad\forall(l,n,m):F_{ln}=T_{lm}=1\\
& \hspace{-2mm} q^f_l \!-\! \tilde{q}^f_l \!=\! \tfrac{\partial q^f_l}{\partial v_n}(v_n\!-\!\tilde{v}_n) \!+\! \tfrac{\partial q^f_l}{\partial v_m}(v_m\!-\!\tilde{v}_m) \!+\! \tfrac{\partial q^f_l}{\partial \theta_{nm}}(\theta_{nm}\!-\!\tilde{\theta}_{nm}), \nonumber \\
&\qquad\forall(l,n,m):F_{ln}=T_{lm}=1\\
& \hspace{-2mm} p^t_l \!-\! \tilde{p}^t_l \!=\! \tfrac{\partial p^t_l}{\partial v_n}(v_n\!-\!\tilde{v}_n) \!+\! \tfrac{\partial p^t_l}{\partial v_m}(v_m\!-\!\tilde{v}_m) \!+\! \tfrac{\partial p^t_l}{\partial \theta_{mn}}(\theta_{mn}\!-\!\tilde{\theta}_{mn}), \nonumber \\
& \qquad\forall(l,n,m):F_{ln}=T_{lm}=1\\
& \hspace{-2mm} q^t_l \!-\! \tilde{q}^t_l \!=\! \tfrac{\partial q^t_l}{\partial v_n}(v_n\!-\!\tilde{v}_n) \!+\! \tfrac{\partial q^t_l}{\partial v_m}(v_m\!-\!\tilde{v}_m) \!+\! \tfrac{\partial q^t_l}{\partial \theta_{mn}}(\theta_{mn}\!-\!\tilde{\theta}_{mn}), \nonumber \\
& \qquad\forall(l,n,m):F_{ln}=T_{lm}=1\\
& \eqref{eq:ac_opf_pglim}-\eqref{eq:ac_opf_stmax} 
\end{align} \label{eq:ac_taylor_opf}
\end{subequations}
\noindent where $\tilde{p}^f_l, \tilde{q}^f_l, \tilde{p}^t_l, \tilde{q}^t_l$ are the power flows of the given operating point obtained through equations \eqref{eq:ac_opf_pf}--\eqref{eq:ac_opf_qt}. Despite involving a greater number of variables and constraints, optimization model \eqref{eq:ac_taylor_opf} retains the same structural advantages as model \eqref{eq:dc_opf}. Its convex objective function and constraints ensure that it can be solved to global optimality in a finite number of steps using efficient algorithms. However, unlike the DC-OPF formulation, model \eqref{eq:ac_taylor_opf} incorporates both reactive power flows and voltage magnitudes, which allows reducing the degree of infeasibility and/or the cost of repair of the so-obtained approximate solutions. Although model \eqref{eq:ac_taylor_opf} is a convex approximation of the true model \eqref{eq:ac_opf}, it retains a similar level of interpretability, as it provides insights into the binding physical constraints and explains why further reductions in dispatch costs are not possible without violating the technical limits of the generating units or network branches. This represents a significant advantage over black-box methods based on neural networks, which rely on complex, non-transparent mappings learned from data, making it challenging to understand how the results are derived or to trace the reasoning behind the solutions.

Although Taylor expansion has previously been used to iteratively solve the AC-OPF problem \cite{burchett1982large, zhang2013relaxed, mhanna2021exact}, this is, to the best of our knowledge, the first time that learning techniques have been employed to compute, based on historical data, the Taylor expansion around an operating point expected to be close to the optimal solution of the AC-OPF. Furthermore, unlike other procedures that use Taylor approximations, our approach is non-iterative, which significantly reduces the overall computational burden. Since the first-order Taylor expansion is more accurate near the operating point where it is applied, the solution of model \eqref{eq:ac_taylor_opf} approximates that of model \eqref{eq:ac_opf} as long as its optimal solution is close enough to the operating point used for the Taylor expansion. Therefore, the primary challenge with this method lies in selecting an appropriate operating point for the Taylor expansion. To address this, we introduce in Section \ref{sec:methodology} a simple, fast, and intuitive learning-based algorithm to infer this operating point, specifically utilizing a $K$-nearest neighbor regression technique \cite{peterson2009k}.

For comparison purposes, we also consider a Taylor expansion of the OPF problem that focuses solely on the dependency of active power flows on voltage angles, while neglecting reactive power flows and voltage magnitudes. This approach yields a model similar to the classical DC-OPF but with a linear relationship between active power flows and voltage angles that varies with the operating point. The simplified Taylor expansion model is formulated as follows:
\begin{subequations}
\begin{align}
& \min_{p_g,p^f_l,p^t_l,\theta_n} \sum_g c_g(p_g) \\
& \text{subject to} \nonumber\\
& \sum_{g\in\mathcal{G}_n} p_g - p^d_n = \tilde{v}_n\tilde{v}_nG^{sh}_n + \sum_{l} (F_{ln}p^f_l + T_{ln}p^t_l), \forall n\\
& p^f_l - \tilde{p}^f_l = \tfrac{\partial p^f_l}{\partial \theta_{nm}}(\theta_{nm}-\tilde{\theta}_{nm}) \label{eq:dc_taylor_opf_pf}\\
& p^t_l - \tilde{p}^t_l =  \tfrac{\partial p^t_l}{\partial \theta_{mn}}(\theta_{mn}-\tilde{\theta}_{mn}) \label{eq:dc_taylor_opf_pt}\\
& \underline{p}_g \leq p_g \leq \overline{p}_g, \forall g \\
& -\overline{s}_l \leq p^f_l \leq \overline{s}_l, \forall l  \\
& -\overline{s}_l \leq p^t_l \leq \overline{s}_l, \forall l  
\end{align} \label{eq:dc_taylor_opf}
\end{subequations}
All the derivatives required for models \eqref{eq:ac_taylor_opf} and \eqref{eq:dc_taylor_opf} are provided in the Appendix.


\section{Learning-based methods}\label{sec:methodology}

In this section, we describe the various learning-based approaches compared in this paper to infer the solution of the AC-OPF problem using nodal active and reactive demands as features. Let us denote $\mathbf{d} = [(p^d_n, q^d_n) \forall n]$ as the vector containing the active and reactive demands at all nodes. We characterize the solution of the optimal power flow by the active power output of the generators $\mathbf{p} = [(p_g) \forall g]$ and the voltage magnitudes at the nodes with generating units $\mathbf{v} = [(v_n) \forall n \in \mathcal{N}_G]$, where $\mathcal{N}_G$ is the subset of nodes with generating units. Using $\mathbf{d}, \mathbf{p}, \mathbf{v}$, we can solve a conventional AC power flow to obtain the remaining variables, including reactive generation levels, voltage magnitudes at load buses, voltage angles in the network, and active and reactive power flows through all branches. 

As discussed in Section \ref{sec:introduction}, finding the global optimal solution for AC-OPF instances is computationally demanding due to the problem's non-convex nature and the absence of convergence guarantees. However, we assume that the historical solutions used in our analysis have been computed offline by solving model \eqref{eq:ac_opf} to global optimality, without constraints on computational resources. In this context, and aligned with common assumptions in the supervised learning literature, this work presumes the availability of a dataset containing historical globally optimal solutions to the AC-OPF model \eqref{eq:ac_opf} denoted by ($\mathbf{p}^*, \mathbf{v}^*$) for different demand profiles ($\mathbf{d}$). Specifically, we have access to $\{(\mathbf{d}_i,\mathbf{p}^*_i,\mathbf{v}^*_i),\forall i \in \mathcal{I}\}$, where $\mathcal{I}$ is the set of training instances.  In practice, when optimal solutions for large networks are unavailable, state estimation outputs can serve as proxies, as these operating points statistically satisfy physical constraints and have near-minimal costs.

We use this training data to infer as accurately as possible the solution of the AC-OPF for a new instance $i'$ with a demand profile $\mathbf{d}_{i'}$. First, we consider a $K-$nearest neighbor (KNN) regression that infers the solution to the AC-OPF problem as the average of the solutions of the nearest neighbors according to a given distance \cite{peterson2009k}. In particular, for a given value of $K$, we determine the subset of instances $\mathcal{I}_K$, with $|\mathcal{I}_K|=K$, that have the lowest Euclidean distance with respect to $\mathbf{d}_{i'}$. Then, the solution is inferred as the average of the closest neighbors, that is, $\hat{\mathbf{p}}_{i'}=(\sum_{i\in\mathcal{I}_K}\mathbf{p}^*_i)/K$ and $\hat{\mathbf{v}}_{i'}=(\sum_{i\in\mathcal{I}_K}\mathbf{v}^*_i)/K$. This is a simple and intuitive learning method that can achieve good accuracy, especially with large training datasets. However, it is purely data-driven and lacks inherent knowledge of the underlying physics governing the problem. This can lead to performance degradation when the input data deviates significantly from the training set, as the closest neighbors may not accurately represent the new instance $i'$. 

At the other end of the spectrum of learning-based approaches are neural networks (NN), which boast high versatility and have demonstrated their ability to capture complex, non-linear relationships between inputs and outputs in optimization problems due to their vast number of parameters \cite{abiodun2018state}. However, this very complexity comes at a cost. Neural networks are often criticized for their lack of interpretability, making it difficult to understand how they arrive at their solutions. Additionally, they require substantial training data and can be challenging to train effectively. In this paper, we employ two distinct neural networks to predict the active power output of generators ($\mathbf{p}$) and the voltage magnitudes at the buses with generation ($\mathbf{v}$). Detailed information about the neural network architecture used for learning these variables of the AC-OPF solution is provided in Section \ref{sec:simulations}.

The learning approach proposed in this paper begins with a KNN regression to estimate an operating point close to the AC-OPF solution for a given demand profile, $\mathbf{d}_{i'}$. This operating point is then used to solve an AC power flow problem, where the active and reactive demand values ($\mathbf{d}_{i'}$) are fixed. The active power generation for all units, except the slack bus, is set to the values estimated by the KNN ($\hat{\mathbf{p}}_{i'}$), and the voltage magnitudes at the generation nodes are also fixed to the KNN estimates ($\hat{\mathbf{v}}_{i'}$). Although the operating point derived from solving the power flow problem might violate certain inequality constraints \eqref{eq:ac_opf_pglim}-\eqref{eq:ac_opf_stmax}, it inherently satisfies the power balance and power flow equality constraints \eqref{eq:ac_opf_balp}-\eqref{eq:ac_opf_qt} of the AC-OPF problem. This property makes it a suitable reference for calculating a first-order Taylor expansion of the nonlinear power flow equations \eqref{eq:ac_opf_pf}--\eqref{eq:ac_opf_qt} around this point. The final step involves solving the convex models \eqref{eq:ac_taylor_opf} or \eqref{eq:dc_taylor_opf} to determine the solution of the optimal power flow problem. This hybrid method combines the strengths of model-based and learning-based approaches, offering a balance between interpretability, computational efficiency, and the ability to capture the physical relationships inherent in the power flow problem. Two variants of the proposed approach are presented: KT-AC and KT-DC. The KT-AC variant solves the convex problem \eqref{eq:ac_taylor_opf}, which accounts for both active and reactive power flows through the branches. In contrast, the KT-DC variant solves the convex problem \eqref{eq:dc_taylor_opf}, focusing solely on the relationship between active power flow and voltage angles, while neglecting reactive power and voltage magnitudes. Algorithm \ref{alg:proposed} includes a pseudo-code of the proposed approach.

\begin{algorithm}[H]
  \caption{Pseudo-code of proposed approach} \label{alg:proposed}
\textbf{Input:} Historical data $\{(\mathbf{d}_i,\mathbf{p}^*_i,\mathbf{v}^*_i),\forall i \in \mathcal{I}\}$, number of neighbors $K$, and test demand profile $\mathbf{d}_{i'}$.

\textbf{Output:} Active power output of generators $\mathbf{p}_{i'}$ \mbox{and voltage magnitudes at generating buses $\mathbf{v}_{i'}$.} 
 \begin{algorithmic}[1]
 \vspace{-4mm}
    \STATE Determine $K$ closest neighbors $\mathcal{I}_K$.
    \STATE Compute $\hat{\mathbf{p}}_{i'}= \frac{\sum_{i\in\mathcal{I}_K}\mathbf{p}^*_i}{K}$ and $\hat{\mathbf{v}}_{i'}= \frac{\sum_{i\in\mathcal{I}_K}\mathbf{v}^*_i}{K}$.
    \STATE Solve power flow with $\mathbf{d}_{i'}$,$\hat{\mathbf{p}}_{i'}$,$\hat{\mathbf{v}}_{i'}$ to obtain $\{\tilde{v}_n, \tilde{\theta}_n, \forall n\}$.
    \STATE Compute derivatives \eqref{eq:derivatives} and solve \eqref{eq:ac_taylor_opf} or \eqref{eq:dc_taylor_opf}.
    \STATE If feasible, return  $\mathbf{p}_{i'},\mathbf{v}_{i'}$. Otherwise, return $\hat{\mathbf{p}}_{i'},\hat{\mathbf{v}}_{i'}$.
\end{algorithmic}
\end{algorithm}

In summary, we compare in this paper the performance of the following six approaches:
\begin{itemize}[left=0mm]
\item AC-OPF: Conventional AC Optimal Power Flow, which provides the most accurate solution.
\item DC-OPF: Conventional DC Optimal Power Flow, a commonly used approximation.
\item KNN: Nearest Neighbor Regression, a straightforward learning-based approach.
\item NN: Neural Network Regression, a more complex learning-based approach.
\item KT-AC: Nearest Neighbor Regression combined with Taylor expansion of active and reactive power flows (proposed).
\item KT-DC: Nearest Neighbor Regression combined with Taylor expansion of active power flow only (proposed).
\end{itemize}

Although the methods mentioned above do not constitute an exhaustive list of all approaches in the literature aimed at solving the AC-OPF problem, we believe they represent a broad spectrum of methodologies, including classical and widely used model-based approximations (e.g., DC-OPF), simple and interpretable learning-based methods (e.g., KNN), and more complex black-box learning methods (e.g., neural networks).

At this point, we must make an important clarification. For any given demand $\mathbf{d}_{i'}$, the DC-OPF, KNN, and NN approaches always provide a corresponding operating point. However, optimization problems \eqref{eq:ac_taylor_opf} or \eqref{eq:dc_taylor_opf} may occasionally become infeasible for certain demands $\mathbf{d}_{i'}$ and the operating point $\{\tilde{v}_n, \tilde{\theta}_n, \forall n\}$ used for the Taylor expansion. In such rare cases, one can address infeasibilities by solving a modified version of problems \eqref{eq:ac_taylor_opf} and \eqref{eq:dc_taylor_opf}, incorporating slack variables to increase the maximum power flow capacities of network branches and adjust the maximum/minimum voltage levels at buses. These slack variables are penalized in the objective function, ensuring that they take non-zero values only when the original models are infeasible. However, given the extremely low number of infeasible instances and in the interest of simplicity, we have chosen in this paper to directly utilize the solutions provided by the KNN approach in those few cases, as they offer similar levels of accuracy to those achieved by the approach with slack variables.

This should not be confused with the general observation that solutions derived from all model- and learning-based approximation methods (DC-OPF, KNN, NN, KT-AC, KT-DC) are typically not AC-feasible, as they often fail to satisfy certain constraints of the original AC-OPF model \eqref{eq:ac_opf}.




\section{Comparison}\label{sec:comparison}

In this section, we describe how we compare and evaluate  the performance of the  methods described in the previous section against the solution of the AC-OPF benchmark. The output of each approach is characterized by the active power generation levels ($\hat{\mathbf{p}}$) and the voltage magnitudes at the buses where these generators are located ($\hat{\mathbf{v}}$), while the true solution provided by the AC-OPF is denoted as $\mathbf{p}^*$ and $\mathbf{v}^*$. Since the DC-OPF model does not account for voltage magnitudes, they are assumed to be 1 p.u. for all buses. Similarly, the KT-DC approach, which is akin to the DC-OPF model \eqref{eq:dc_taylor_opf}, also omits voltage magnitudes. For this method, though, the voltage magnitudes at generating nodes are set to the values provided by the KNN regression that it uses. 

We initially evaluate the performance of the different methods using the mean squared error (MSE) of the active power generation levels and voltage magnitudes at the generation buses. However, since a lower MSE does not necessarily imply lower suboptimality, we also assess the methods using additional metrics that quantify the infeasibility and suboptimality of the obtained solutions. For each of the five methods, we compute the following three metrics:

\begin{itemize}[left=0mm]
\item Mean Squared Error (MSE): This metric is purely statistical and measures the mean squared error of the active power output levels of the generators ($\hat{\mathbf{p}}$) and the voltage magnitudes at the nodes with generation ($\hat{\mathbf{v}}$). This is, for example, the performance metric used in reference \cite{joswig2022opf}. We denote these measures as $MSE(\mathbf{p})$ and $MSE(\mathbf{v})$, respectively, and compute them as follows:
\begin{equation*}
MSE(\mathbf{p}) = \|\hat{\mathbf{p}}-\mathbf{p}^*\|^2_2 \qquad  MSE(\mathbf{v}) = \|\hat{\mathbf{v}}-\mathbf{v}^*\|^2_2   
\end{equation*}
\item Infeasibility: This metric assesses the infeasibility of the obtained solutions. For each demand profile, the active and reactive loads at the demand nodes are fixed. The active power outputs of all generators, except for the slack unit, are set to the values obtained from each approach. Similarly, the voltage magnitudes are fixed to the values provided by the solution. Then, a standard AC power flow calculation is performed, which involves solving a system of 2$N$ nonlinear equations with 2$N$ unknowns, to identify an operating point that satisfies both the power balance and power flow equations. We then assess the average normalized violations of the inequality constraints, following the methodology outlined in \cite{venzke2020inexact}.
\begin{equation*}
    VIOL(\mathbf{x}) = \tfrac{1}{|\mathcal{J}|} \sum_{j \in \mathcal{J}} \tfrac{\max(x_j-\overline{x}_j,\underline{x}_j-x_j,0)}{\overline{x}_j-\underline{x}_j} \times 100\,\%
\end{equation*}
\noindent where $\mathbf{x}$ is a vector variable representing the voltage magnitudes, power generation levels, or power flows through the branches. The index $j$ refers to buses, generating units, or network branches, depending on the specific variable $\mathbf{x}$. The terms $\overline{x}_j$ and $\underline{x}_j$ denote the upper and lower bounds for the variable $\mathbf{x}$, respectively. In particular, in our analysis we consider the following average constraints violations:
\begin{itemize}[left=0mm]
    \item $VIOL(\mathbf{v})$: constraint violation of voltage magnitude.
    \item $VIOL(\mathbf{p})$: constraint violation of active power.
    \item $VIOL(\mathbf{q})$: constraint violation of reactive power.
    \item $VIOL(\mathbf{s})$: constraint violation of apparent power.
\end{itemize}
It is important to note that constraint violations can have different interpretations depending on the approach used. For instance, the solutions $\hat{\mathbf{p}}$ obtained from KT-DC and KT-AC are derived from optimization problems that include constraints on the active power output of generators. Thus, a $VIOL(\mathbf{p})$ value greater than 0 indicates that the slack generation levels are outside the valid range. Conversely, the solution $\hat{\mathbf{p}}$ produced by a neural network might violate the technical limits of any generating unit. Similarly, $VIOL(\mathbf{v})$ refers to the voltage magnitude violation at all network nodes. While the voltage magnitudes at generation nodes in the KT-AC approach consistently remain within their specified limits due to enforced constraints, the voltage magnitudes at demand buses, as determined by the power flow solution, may still exceed these bounds.

\item Redispatch: This metric evaluates the redispatch actions (repair) required to achieve an AC-feasible operating point as close as possible to the one obtained by each approach. To do this, we solve the optimization problem \eqref{eq:ac_feas}, which minimizes the slack variables of the active power output of generators ($\Delta\mathbf{p}$) and the voltage magnitude levels ($\Delta\mathbf{v}$) to find an operating point that satisfies the constraints of the AC-OPF. For any solution provided by the methods described in Section \ref{sec:methodology}, characterized by $\hat{\mathbf{p}}$ and $\hat{\mathbf{v}}$, the following feasibility problem is solved:
\begin{subequations}
\begin{align}
& \min_{\Delta p_g,\Delta v_n} \sum_g |\Delta p_g| + \sum_{n \in \mathcal{N}_G} |\Delta v_n| \label{eq:ac_feas_of}\\
& \text{subject to} \nonumber\\
& \eqref{eq:ac_opf_balp}-\eqref{eq:ac_opf_stmax} \\
& p_g = \hat{p}_g + \Delta p_g, \forall g \\
& v_n = \hat{v}_n + \Delta v_n, \forall n \in \mathcal{N}_G
\end{align} \label{eq:ac_feas}
\end{subequations}
Note that model \eqref{eq:ac_feas} is just a norm-1 projection into the AC-feasible space of the AC-OPF problem. In practice, if the slack variables are sufficiently small, these infeasibilities can often be accomodated by the system through minor violations of line capacity limits or slight deviations in system frequency. However, if the slack variables exceed acceptable thresholds, redispatch actions may be necessary to restore system feasibility and maintain operational reliability.

Additionally, we provide the increase in operating cost of this AC-feasible point compared to the solution of the AC-OPF, as a measure of suboptimality, which we denote as $\Delta C$ and compute as:
\begin{equation*}
\Delta C = \tfrac{\sum_g c_g(\hat{p}_g + \Delta p_g) - \sum_gc_g(p^*_g)}{\sum_g c_g(p^*_g)} \times 100\,\%    
\end{equation*}
\end{itemize}

\section{Numerical simulations}\label{sec:simulations}

\begin{table*}
\centering
\caption{Results with $8\,000$ training data set} \label{tab:results8000}
\begin{tabular}{lcccccccccc}
\toprule
  System &       Approach &  $MSE(\mathbf{p})$ & $MSE(\mathbf{v})$ &  $VIOL(\mathbf{p})$ &  $VIOL(\mathbf{q})$ &  $VIOL(\mathbf{s})$ &  $VIOL(\mathbf{v})$ &  $\Delta\mathbf{p}$ &  $\Delta\mathbf{v}$ &  $\Delta C$ \\
\midrule
  case14 &         DC-OPF &  1.8e-02 &  1.7e-03 &  1.3e-01 &  1.2e+02 &  \textbf{0.0e+00} &  2.9e-01 &  6.2e-02 &  2.3e-02 &  6.5e+00 \\
   case14 &        KNN &  1.5e-02 &  1.4e-04 &  2.8e-02 &  4.4e+00 &  \textbf{0.0e+00} &  2.1e-03 &  4.4e-02 &  3.8e-03 &  5.6e+00 \\
   case14 &         NN &  5.0e-04 &  1.1e-05 &  7.6e-03 &  4.8e+00 &  \textbf{0.0e+00} &  5.8e-02 &  4.6e-03 &  1.1e-03 &  5.6e-01 \\
  case14 &  KT-DC &  2.1e-02 &  1.4e-04 &  \textbf{2.0e-04} &  5.0e+00 &  \textbf{0.0e+00} &  2.1e-03 &  5.5e-02 &  5.5e-03 &  1.1e-01 \\
  case14 &  KT-AC &  \textbf{1.0e-04} &  \textbf{3.0e-06} &  3.0e-04 &  \textbf{2.5e-01} &  \textbf{0.0e+00} &  \textbf{1.0e-04} &  \textbf{4.2e-03} &  \textbf{2.0e-04} &  \textbf{2.0e-02} \\  
  \midrule
  case30 &         DC-OPF &  9.3e-03 &  2.1e-03 &  1.7e-03 &  8.6e+01 &  3.1e-01 &  2.7e+00 &  3.6e-02 &  3.4e-02 &  3.5e+00 \\
    case30 &        KNN &  9.1e-03 &  8.7e-05 &  1.1e-03 &  1.5e+00 &  1.4e-01 &  2.9e-02 &  3.4e-02 &  3.0e-03 &  3.7e+00 \\
      case30 &         NN &  4.3e-03 &  \textbf{1.2e-05} &  5.7e-03 &  1.3e+00 &  6.5e-03 &  9.5e-02 &  1.0e-02 &  1.3e-03 &  1.6e+00 \\
  case30 &  KT-DC &  6.1e-03 &  8.3e-05 &  \textbf{0.0e+00} &  1.8e+00 &  2.3e-02 &  3.0e-02 &  1.3e-02 &  3.1e-03 &  \textbf{7.0e-02} \\

  case30 &  KT-AC &  \textbf{0.0e+00} &  9.2e-05 &  \textbf{0.0e+00} &  \textbf{8.8e-02} &  \textbf{2.9e-03} &  \textbf{3.0e-04} &  \textbf{9.0e-04} &  \textbf{1.0e-04} &  2.4e-01 \\

  \midrule
  case57 &         DC-OPF &  3.9e-01 &  1.3e-03 &  3.3e+00 &  2.2e+02 &  \textbf{0.0e+00} &  1.9e+00 &  2.4e-01 &  2.6e-02 &  2.4e-01 \\
    case57 &        KNN &  1.9e-01 &  1.2e-04 &  1.8e+00 &  1.6e+01 &  \textbf{0.0e+00} &  1.3e-01 &  1.3e-01 &  5.8e-03 &  4.3e-01 \\
      case57 &         NN &  1.7e-01 &  4.5e-05 &  5.6e-01 &  1.1e+01 &  \textbf{0.0e+00} &  6.1e-02 &  9.2e-02 &  3.3e-03 &  5.1e-01 \\
  case57 &  KT-DC &  4.9e-01 &  1.2e-04 &  2.3e-01 &  3.9e+01 &  1.0e-04 &  2.3e-01 &  2.4e-01 &  7.7e-03 &  1.5e-01 \\

  case57 &  KT-AC &  \textbf{4.2e-02} &  \textbf{3.2e-05} &  \textbf{1.4e-01} &  \textbf{3.3e+00} &  \textbf{0.0e+00} &  \textbf{4.5e-03} &  \textbf{3.1e-02} &  \textbf{8.0e-04} &  \textbf{7.0e-02} \\

  \midrule
 case118 &         DC-OPF &  1.2e-01 &  1.7e-03 &  \textbf{0.0e+00} &  7.6e+01 &  2.6e-01 &  3.3e-02 &  4.1e-02 &  1.4e-02 &  7.0e+00 \\
  case118 &        KNN &  1.2e-01 &  2.6e-04 &  1.5e-02 &  2.4e+01 &  2.1e-01 &  2.2e-03 &  4.4e-02 &  4.3e-03 &  5.6e+00 \\
   case118 &         NN &  7.7e-02 &  \textbf{2.0e-04} &  2.3e-01 &  2.4e+01 &  1.0e-01 &  1.7e-01 &  3.7e-02 &  4.4e-03 &  4.8e+00 \\
 case118 &  KT-DC &  7.0e-02 &  2.5e-04 &  \textbf{0.0e+00} &  2.5e+01 &  \textbf{1.3e-02} &  2.0e-03 &  \textbf{9.5e-03} &  4.7e-03 &  \textbf{1.2e+00} \\

 case118 &  KT-AC &  \textbf{5.1e-02} &  9.1e-04 &  \textbf{0.0e+00} &  \textbf{2.3e+00} &  3.0e-02 &  \textbf{9.0e-04} &  1.0e-02 &  \textbf{1.3e-03} &  2.2e+00 \\
\bottomrule
\end{tabular}
\end{table*}

In this section, we compare the methods described in Section \ref{sec:methodology} using networks of different sizes. As discussed in \cite{joswig2022opf}, the data-driven methods proposed in the literature to learn the solution of the AC-OPF problem are typically benchmarked using scenarios with low variability. Typically, the demand profile instances are generated within a specific percentage range (5\%, 10\%, 20\%) around the nominal value. As shown in \cite{joswig2022opf}, these datasets are not particularly challenging, and conclusions drawn from such analyses can be biased. To address this, the authors of \cite{joswig2022opf} propose a methodology to create more challenging datasets of AC-OPF instances, which are better suited for comparing the performance of different data-driven methodologies. We directly use the datasets created by these authors for four different networks of 14, 30, 57, and 118 buses. These datasets can be downloaded from \cite{Joswig-Jones2021}. Each dataset includes $10\,000$ instances, split into $8\,000$ training instances and $2\,000$ test instances. All the results provided in this section are average results over the $2\,000$ test instances.

The non-convex AC-OPF model \eqref{eq:ac_opf} and the non-convex redispatch model \eqref{eq:ac_feas} are both solved using IPOPT 3.12.8 with the flat voltage profile as the initial solution. The convex models \eqref{eq:dc_opf}, \eqref{eq:ac_taylor_opf}, and \eqref{eq:dc_taylor_opf} are solved using Gurobi 9.0.1. For all the approaches that make use of nearest neighbor regression (KNN, KT-DC, KT-AC), the value of $K$ is fixed at 100. Preliminary results indicate that variations in the parameter, ranging from 5 to 500, have only a slight impact on the performance metrics of these methods. Consequently, a detailed sensitivity analysis of $K$ is not included in this paper. For the NN approach we use two neural networks: one to learn the active power output and another to learn the voltage magnitudes at the generation nodes. We utilize the same neural network architecture proposed in \cite{joswig2022opf}, which includes three hidden layers, with the width of the first two being equal to the number of inputs (i.e., the size of vector $\mathbf{d}$), while the width of the last hidden layer equals the number of outputs (i.e., the size of vector $\mathbf{p}$ or $\mathbf{v}$). We use the ReLU activation function between all layers. We adjust the parameters of the neural network using TensorFlow 2.13.0 with 100 epochs and a batch size of 32. As previously discussed, neural networks have numerous hyperparameters that can be challenging to tune properly. Additionally, the initial values of the layer parameters can affect the results, making it difficult to reproduce the findings in \cite{joswig2022opf}. To address this, we train each neural network 100 times with different random seeds and select the one that provides the highest accuracy. By doing so, our results closely match those reported in \cite{joswig2022opf}.

Table~\ref{tab:results8000} presents a comprehensive overview of the performance metrics, as detailed in Section \ref{sec:comparison}, averaged across $2\,000$ test instances for each network size. In this table, the metrics $MSE(\mathbf{p})$, $MSE(\mathbf{v})$, $\Delta\mathbf{p}$, and $\Delta\mathbf{v}$ are presented in per unit (p.u.), while the constraint violations and $\Delta C$ are expressed as percentages. Besides, $\Delta\mathbf{p}$ and $\Delta\mathbf{v}$ are average values over the number of generators and  the number of buses with generation, respectively. To ease the interpretation of the results, the best performance that is achieved for each metric and network size is highlighted in bold. 

While the DC-OPF approach remains a popular approximation method, our results demonstrate its inferior performance across most evaluation criteria. Specifically, the solutions from DC-OPF incur average cost increases of 6.5\%, 3.5\%, 0.24\%, and 7\% for 14-, 30-, 57-, and 118-node networks, respectively. Furthermore, the method exhibits significant violations of reactive generation constraints, averaging 120\%, 86\%, 220\%, and 76\% for the same networks. These findings underline the limitations of the DC-OPF, despite its widespread use. Interestingly, KT-DC demonstrates a substantial improvement over DC-OPF, despite the fact that models \eqref{eq:dc_opf} and \eqref{eq:dc_taylor_opf} are of equivalent size and complexity. This improvement is attributed to a more refined linear approximation of active power flows with respect to angle differences. Unlike the static linear relationship assumed in DC-OPF, the proposed KT-DC adjusts the intercept and slope of the linear approximation based on the specific operating point, leading to enhanced accuracy. As a result, the average cost increase for KT-DC is significantly reduced to 0.11\%, 0.07\%, 0.15\%, and 1.2\% for 14-, 30-, 57-, and 118-node networks, respectively. This indicates a much closer approximation of active generation levels to the AC-OPF benchmark. Although the KT-DC method overlooks reactive power flows, the reactive generation levels determined in the subsequent power flow calculations align more closely with the AC-OPF values. This alignment significantly reduces average reactive generation constraint violations to 5\%, 1.8\%, 39\%, and 25\% for the respective network sizes.

A comparison of the learning-based approaches shows that while NN generally surpasses KNN across the four systems, the differences are often minor. Particularly, for the 30- and 118-bus networks, KNN even yields lower values for $VIOL(\mathbf{p})$ and $VIOL(\mathbf{v})$. Therefore, although KNN may result in a reduction in performance due to its simplicity, it offers significant advantages, including lower data and computational requirements, as well as greater interpretability and explainability of the results. Furthermore, the low MSE values for the KNN approach validate the use of this solution as the operating point for the first-order Taylor expansion.

Finally, comparing KT-DC and KT-AC reveals that the latter generally performs better, especially in approximating reactive power generation. While KT-AC also includes a linear formulation of the reactive power flow equations, it accounts for both reactive flows and voltage magnitudes, resulting in more accurate outcomes. Despite this, KT-AC retains the advantages of a convex problem, offering convergence and optimality guarantees while being as efficiently solvable as the conventional DC-OPF problem. Additionally, the results show that KT-AC outperforms NN in most performance metrics across all network sizes. The only exceptions are the MSE of voltage magnitudes for the 30- and 118-bus networks, where NN performs slightly better. In all other cases, KT-AC delivers significantly superior performance, with improvements sometimes reaching one or two orders of magnitude. Thus, apart from being much simpler, more interpretable, and easier to train than neural networks, the proposed KT-AC provides solutions that more accurately approximate the true solution of the AC-OPF problem. 

As noted in Section \ref{sec:methodology}, there are instances where the optimization models \eqref{eq:ac_taylor_opf} or \eqref{eq:dc_taylor_opf} may become infeasible. In such cases, the solution from KT-AC and KT-DC defaults to the result provided by the KNN regression. Table \ref{tab:infes} displays the percentage of infeasible cases among the $2\,000$ test instances for each network size. It is observed that the incidence of infeasible cases is relatively low, especially for KT-AC.

\begin{table}[]
\centering
\caption{Percentage of test instances for which models \eqref{eq:ac_taylor_opf} and \eqref{eq:dc_taylor_opf} are infeasible for the KT-AC and KT-DC approaches, respectively.}
\begin{tabular}{lcccc}
\toprule
 & case14 & case30 & case57& case118 \\
\midrule
\# Infes KT-DC  & 0.00\% & 0.85\% & 0.00\%& 2.40\% \\
\# Infes KT-AC  & 0.00\% & 0.65\% & 0.05\%& 0.00\% \\
\bottomrule
\end{tabular}
\label{tab:infes}
\end{table}

The data-driven approaches compared in this study may be influenced by the size of the training dataset. To investigate this, Table \ref{tab:results2000} presents the computational results obtained when the training sample size is reduced from $8\,000$ to $2\,000$ samples. Comparing these results with those in Table \ref{tab:results8000}, we observe that the performance of KNN and KT-AC has remained relatively stable, whereas NN shows a significant degradation in performance. Specifically, for the reduced training set, KT-AC consistently outperforms NN across all analyzed metrics and network configurations. This highlights the superior data efficiency of the proposed learning-based method since KT-AC not only offers advantages in terms of speed, simplicity, and interpretability but also larger resilience to data scarcity, making it more practical for real-world applications.

\begin{table*}
\centering
\caption{Results with $2\,000$ training data set} \label{tab:results2000}
\begin{tabular}{lcccccccccc}
\toprule
  System &       Approach &  $MSE(\mathbf{p})$ & $MSE(\mathbf{v})$ &  $VIOL(\mathbf{p})$ &  $VIOL(\mathbf{q})$ &  $VIOL(\mathbf{s})$ &  $VIOL(\mathbf{v})$ &  $\Delta\mathbf{p}$ &  $\Delta\mathbf{v}$ &  $\Delta C$ \\
\midrule  
  case14 &        KNN &  1.7e-02 &  1.5e-04 &  3.5e-02 &     5.0e+00 &  \textbf{0.0e+00} &  2.3e-03 &  4.9e-02 &  4.0e-03 &  6.3e+00 \\
   case14 &         NN &  1.6e-03 &  1.5e-04 &  3.4e-02 &    2.1e+01 &  \textbf{0.0e+00} &  3.9e-01 &  1.1e-02 &  5.8e-03 &  1.2e+00 \\
  case14 &  KT-AC &  \textbf{2.0e-04} &  \textbf{4.0e-06} &  \textbf{4.0e-04} &     \textbf{3.0e-01} &  \textbf{0.0e+00} &  \textbf{1.0e-04} &  \textbf{4.6e-03} &  \textbf{3.0e-04} &  \textbf{3.0e-02} \\
 
  \midrule   
  case30 &        KNN &  1.1e-02 &  9.3e-05 &  9.3e-03 &     2.0e+00 &  1.9e-01 &  2.8e-02 &  4.2e-02 &  3.2e-03 &  4.2e+00 \\
    case30 &         NN &  7.7e-03 &  2.4e-04 &  3.7e-02 &     9.7e+00 &  4.5e-02 &  6.2e-01 &  2.0e-02 &  6.1e-03 &  3.7e+00 \\
  case30 &  KT-AC &  \textbf{1.0e-04} &  \textbf{9.2e-05} &  \textbf{0.0e+00} &    \textbf{ 1.0e-01} &  \textbf{3.2e-03} &  \textbf{3.0e-04} &  \textbf{1.0e-03} &  \textbf{1.0e-04} & \textbf{ 2.5e-01} \\

  \midrule   
  case57 &        KNN &  2.1e-01 &  1.4e-04 &  2.7e+00 &    2.0e+01 &  \textbf{0.0e+00} &  1.3e-01 &  1.4e-01 &  6.4e-03 &  4.6e-01 \\
   case57 &         NN &  3.5e-01 &  8.1e-03 &  1.9e+00 &    9.3e+01 &  7.0e-04 &  3.4e+01 &  2.1e-01 &  6.8e-02 &  1.5e+00 \\ 
  case57 &  KT-AC &  \textbf{5.1e-02} &  \textbf{3.4e-05} & \textbf{ 1.9e-01} &     \textbf{4.4e+00} &  \textbf{0.0e+00} & \textbf{ 6.2e-03} &  \textbf{3.8e-02 }&  \textbf{9.0e-04} &  \textbf{9.0e-02} \\
 
  \midrule   
 case118 &        KNN &  1.1e-01 &  \textbf{2.6e-04} &  2.5e-02 &    2.6e+01 &  2.7e-01 &  8.1e-03 &  4.4e-02 &  5.2e-03 &  4.5e+00 \\
  case118 &         NN &  1.2e-01 &  4.4e-02 &  1.7e+01 &   3.9e+02 &  3.2e+01 &  1.9e+01 &  6.4e-02 &  7.1e-02 &  6.2e+00 \\
 case118 &  KT-AC &  \textbf{5.9e-02} &  6.9e-04 &  \textbf{0.0e+00} &     \textbf{2.3e+00} &  \textbf{2.3e-02} &  \textbf{7.0e-04} &  \textbf{9.9e-03} &  \textbf{1.3e-03} &  \textbf{1.7e+00} \\
\bottomrule
\end{tabular}
\end{table*}

To conclude this case study, we examine the potential of using solutions from learning-based approaches as starting points to reduce the number of iterations in the conventional AC-OPF model. Table \ref{tab:iterations} reports the average number of iterations for $2\,000$ test samples of the 118-bus system, considering training datasets of either $8\,000$ or $2\,000$ samples for the learning approaches KNN, NN, and KT-AC. Additionally, we compare the iterations when the starting point is either the direct (and likely AC-infeasible) solution of each learning method or the corresponding feasible projection obtained through the post-processing model \eqref{eq:ac_feas}. With an average of 26.3 iterations for the AC-OPF model, the results reveal the following: for the larger training dataset, all learning methods reduce the average number of iterations. However, for the smaller training dataset, directly using NN solutions significantly increases iterations, highlighting the critical dependence of this approach on high-quality data. Besides, processing learning-based solutions to ensure feasibility consistently reduces iterations compared to using raw outputs. Importantly, these results show that the proposed KT-AC approach achieves the lowest average number of iterations among all methods.

\begin{table}[]
\centering
\caption{Effect of warm start on the average number of iterations of the AC-OPF model for the 118-bus system}
\begin{tabular}{lcccc}
\toprule
& \multicolumn{2}{c}{$8\,000$ training} & \multicolumn{2}{c}{$2\,000$ training} \\
 & INFES & FEAS & INFES & FEAS \\
\midrule
KNN & 22.5 & 21.7 & 22.0 & 20.9 \\
NN & 22.1 & 21.3 & 517.3 & 22.4 \\
KT-AC & \textbf{20.8} & \textbf{20.4} & \textbf{20.8} & \textbf{20.3} \\
\bottomrule
\end{tabular}
\label{tab:iterations}
\end{table}

\section{Conclusions} \label{sec:conclusions}

Currently, power system operators rely on model-based simplifications of the AC power flow equations that are straightforward and interpretable, offering unique solutions that can be reliably obtained using optimization algorithms, albeit often with significant inaccuracies. On the other hand, the scientific community frequently advocates for versatile and powerful approaches, such as neural networks, which provide more accurate solutions in milliseconds. However, these methods come with trade-offs, including reduced parsimony, simplicity and interpretability, as well as reliance on extensive training data, which contribute to operators' reluctance to implement these methods in actual networks. 

In this paper, we argue that while learning-based approaches can help improve the accuracy of power system solutions, it is essential to assign them a role compatible with a sector as critical and fundamental as the electrical industry. Only by recognizing and assuming the specificities of the operation of power systems, and the risk-averse mindset of their operators, we will be able to take the intermediate steps necessary for the seamless integration of machine learning and AI into existing power system operational routines. To support this point, we propose a hybrid strategy that effectively bridges the gap between model-based and learning-based methods. This strategy combines nearest neighbor regression with Taylor expansion, maintaining the simplicity of solving a convex optimization problem, similar to current model-based methods, while also improving accuracy by incorporating historical data. We introduce two versions of our proposed methodology. The first version, akin to the DC-OPF, focuses solely on the linear relationship between active power flows and voltage angles, omitting reactive power and voltage magnitudes. The second version, more aligned with the AC-OPF, incorporates both active and reactive power flows. The proposed method bridges the gap between traditional white-box approximations, such as DC-OPF, and modern black-box neural networks by providing a practical, interpretable, and accurate solution to the AC-OPF problem. Rather than relying entirely on black-box learning-based approaches to directly solve the AC-OPF, our method leverages historical data and learning techniques in a complementary role, preserving the interpretability and transparency of the resulting solution.

Numerical simulations assessing the performance of the proposed methodologies across four power systems of different sizes reveal that they surpass the traditional DC approximation by adapting the linearization to the specific operating conditions. The results also demonstrate that the model incorporating reactive power consistently outperforms its counterpart and achieves accuracy levels on par with neural network models while maintaining superior simplicity and interpretability. Besides, this approach is especially advantageous when training data is scarce, addressing a significant limitation of many learning-based methods. Therefore, our methodology provides a balanced solution that effectively retains the transparency of model-based techniques, but boosted with easily interpretable data-driven methods. We believe this combination can pave the way for a smooth integration of learning-based methods into power system operations, with the potential to build greater trust and confidence among system operators.

Exploring the extension of the proposed approach to scenarios without historical optimal solution data, through the use of self-supervised methodologies, is identified as a promising avenue for future research. Another direction for future research is to extend the proposed algorithm to generate a distribution of dispatch decisions. This can be achieved by performing a Taylor expansion for each of the closest neighbors and solving multiple convex optimization problems, one for each neighbor. 

\bibliographystyle{IEEEtran}
\bibliography{references}

\begin{thebibliography}{10}
\providecommand{\url}[1]{#1}
\csname url@samestyle\endcsname
\providecommand{\newblock}{\relax}
\providecommand{\bibinfo}[2]{#2}
\providecommand{\BIBentrySTDinterwordspacing}{\spaceskip=0pt\relax}
\providecommand{\BIBentryALTinterwordstretchfactor}{4}
\providecommand{\BIBentryALTinterwordspacing}{\spaceskip=\fontdimen2\font plus
\BIBentryALTinterwordstretchfactor\fontdimen3\font minus \fontdimen4\font\relax}
\providecommand{\BIBforeignlanguage}[2]{{%
\expandafter\ifx\csname l@#1\endcsname\relax
\typeout{** WARNING: IEEEtran.bst: No hyphenation pattern has been}%
\typeout{** loaded for the language `#1'. Using the pattern for}%
\typeout{** the default language instead.}%
\else
\language=\csname l@#1\endcsname
\fi
#2}}
\providecommand{\BIBdecl}{\relax}
\BIBdecl

\bibitem{nair2022computational}
A.~S. Nair, S.~Abhyankar, S.~Peles, and P.~Ranganathan, ``Computational and numerical analysis of ac optimal power flow formulations on large-scale power grids,'' \emph{Electric Power Systems Research}, vol. 202, p. 107594, 2022.

\bibitem{frank2016introduction}
S.~Frank and S.~Rebennack, ``An introduction to optimal power flow: Theory, formulation, and examples,'' \emph{IIE transactions}, vol.~48, no.~12, pp. 1172--1197, 2016.

\bibitem{baker2021solutions}
K.~Baker, ``Solutions of dc opf are never ac feasible,'' in \emph{Proceedings of the Twelfth ACM International Conference on Future Energy Systems}, 2021, pp. 264--268.

\bibitem{yang2017solving}
Z.~Yang, H.~Zhong, Q.~Xia, and C.~Kang, ``Solving opf using linear approximations: fundamental analysis and numerical demonstration,'' \emph{IET Generation, Transmission \& Distribution}, vol.~11, no.~17, pp. 4115--4125, 2017.

\bibitem{yang2017linearized}
Z.~Yang, H.~Zhong, A.~Bose, T.~Zheng, Q.~Xia, and C.~Kang, ``A linearized opf model with reactive power and voltage magnitude: A pathway to improve the mw-only dc opf,'' \emph{IEEE Transactions on Power Systems}, vol.~33, no.~2, pp. 1734--1745, 2017.

\bibitem{misra2018optimal}
S.~Misra, D.~K. Molzahn, and K.~Dvijotham, ``Optimal adaptive linearizations of the ac power flow equations,'' in \emph{2018 Power Systems Computation Conference (PSCC)}.\hskip 1em plus 0.5em minus 0.4em\relax IEEE, 2018, pp. 1--7.

\bibitem{low2014convex}
S.~H. Low, ``Convex relaxation of optimal power flow—part i: Formulations and equivalence,'' \emph{IEEE Transactions on Control of Network Systems}, vol.~1, no.~1, pp. 15--27, 2014.

\bibitem{khaloie2024review}
H.~Khaloie, M.~Dolanyi, J.-F. Toubeau, and F.~Vall{\'e}e, ``Review of machine learning techniques for optimal power flow,'' \emph{Available at SSRN 4681955}, 2024.

\bibitem{pan2020deepopf}
X.~Pan, T.~Zhao, M.~Chen, and S.~Zhang, ``Deepopf: A deep neural network approach for security-constrained dc optimal power flow,'' \emph{IEEE Transactions on Power Systems}, vol.~36, no.~3, pp. 1725--1735, 2020.

\bibitem{zamzam2020learning}
A.~S. Zamzam and K.~Baker, ``Learning optimal solutions for extremely fast ac optimal power flow,'' in \emph{2020 IEEE international conference on communications, control, and computing technologies for smart grids (SmartGridComm)}.\hskip 1em plus 0.5em minus 0.4em\relax IEEE, 2020, pp. 1--6.

\bibitem{singh2021learning}
M.~K. Singh, V.~Kekatos, and G.~B. Giannakis, ``Learning to solve the ac-opf using sensitivity-informed deep neural networks,'' \emph{IEEE Transactions on Power Systems}, vol.~37, no.~4, pp. 2833--2846, 2021.

\bibitem{owerko2020optimal}
D.~Owerko, F.~Gama, and A.~Ribeiro, ``Optimal power flow using graph neural networks,'' in \emph{ICASSP 2020-2020 IEEE International Conference on Acoustics, Speech and Signal Processing (ICASSP)}.\hskip 1em plus 0.5em minus 0.4em\relax IEEE, 2020, pp. 5930--5934.

\bibitem{lopez2024optimal}
T.~B. Lopez-Garcia and J.~A. Dom{\'\i}nguez-Navarro, ``Optimal power flow with physics-informed typed graph neural networks,'' \emph{IEEE Transactions on Power Systems}, 2024.

\bibitem{lei2020data}
X.~Lei, Z.~Yang, J.~Yu, J.~Zhao, Q.~Gao, and H.~Yu, ``Data-driven optimal power flow: A physics-informed machine learning approach,'' \emph{IEEE Transactions on Power Systems}, vol.~36, no.~1, pp. 346--354, 2020.

\bibitem{bienstock2022mathematical}
D.~Bienstock, M.~Escobar, C.~Gentile, and L.~Liberti, ``Mathematical programming formulations for the alternating current optimal power flow problem,'' \emph{Annals of Operations Research}, vol. 314, no.~1, pp. 277--315, 2022.

\bibitem{stott2009dc}
B.~Stott, J.~Jardim, and O.~Alsa{\c{c}}, ``Dc power flow revisited,'' \emph{IEEE Transactions on Power Systems}, vol.~24, no.~3, pp. 1290--1300, 2009.

\bibitem{shao2021data}
Z.~Shao, Q.~Zhai, J.~Wu, and X.~Guan, ``Data based linear power flow model: Investigation of a least-squares based approximation,'' \emph{IEEE Transactions on Power Systems}, vol.~36, no.~5, pp. 4246--4258, 2021.

\bibitem{taheri2024optimizing}
B.~Taheri and D.~K. Molzahn, ``Optimizing parameters of the dc power flow,'' \emph{Electric Power Systems Research}, vol. 235, p. 110719, 2024.

\bibitem{li2022numerical}
M.~Li, Y.~Du, J.~Mohammadi, C.~Crozier, K.~Baker, and S.~Kar, ``Numerical comparisons of linear power flow approximations: Optimality, feasibility, and computation time,'' in \emph{2022 IEEE Power \& Energy Society General Meeting (PESGM)}.\hskip 1em plus 0.5em minus 0.4em\relax IEEE, 2022, pp. 1--5.

\bibitem{coffrin2014linear}
C.~Coffrin and P.~Van~Hentenryck, ``A linear-programming approximation of ac power flows,'' \emph{INFORMS Journal on Computing}, vol.~26, no.~4, pp. 718--734, 2014.

\bibitem{yang2018general}
Z.~Yang, K.~Xie, J.~Yu, H.~Zhong, N.~Zhang, and Q.~Xia, ``A general formulation of linear power flow models: Basic theory and error analysis,'' \emph{IEEE Transactions on Power Systems}, vol.~34, no.~2, pp. 1315--1324, 2018.

\bibitem{burchett1982large}
R.~Burchett, H.~Happ, and K.~Wirgau, ``Large scale optimal power flow,'' \emph{IEEE Transactions on Power Apparatus and Systems}, no.~10, pp. 3722--3732, 1982.

\bibitem{zhang2013relaxed}
H.~Zhang, V.~Vittal, G.~T. Heydt, and J.~Quintero, ``A relaxed ac optimal power flow model based on a taylor series,'' in \emph{2013 IEEE Innovative Smart Grid Technologies-Asia (ISGT Asia)}.\hskip 1em plus 0.5em minus 0.4em\relax IEEE, 2013, pp. 1--5.

\bibitem{mhanna2021exact}
S.~Mhanna and P.~Mancarella, ``An exact sequential linear programming algorithm for the optimal power flow problem,'' \emph{IEEE Transactions on Power Systems}, vol.~37, no.~1, pp. 666--679, 2021.

\bibitem{peterson2009k}
L.~E. Peterson, ``K-nearest neighbor,'' \emph{Scholarpedia}, vol.~4, no.~2, p. 1883, 2009.

\bibitem{abiodun2018state}
O.~I. Abiodun, A.~Jantan, A.~E. Omolara, K.~V. Dada, N.~A. Mohamed, and H.~Arshad, ``State-of-the-art in artificial neural network applications: A survey,'' \emph{Heliyon}, vol.~4, no.~11, 2018.

\bibitem{joswig2022opf}
T.~Joswig-Jones, K.~Baker, and A.~S. Zamzam, ``Opf-learn: An open-source framework for creating representative ac optimal power flow datasets,'' in \emph{2022 IEEE Power \& Energy Society Innovative Smart Grid Technologies Conference (ISGT)}.\hskip 1em plus 0.5em minus 0.4em\relax IEEE, 2022, pp. 1--5.

\bibitem{venzke2020inexact}
A.~Venzke, S.~Chatzivasileiadis, and D.~K. Molzahn, ``Inexact convex relaxations for ac optimal power flow: Towards ac feasibility,'' \emph{Electric Power Systems Research}, vol. 187, p. 106480, 2020.

\bibitem{Joswig-Jones2021}
N.~Joswig-Jones, T.~Trager, A.~Zamzam, and K.~Baker, ``Opflearndata: Dataset for learning ac optimal power flow,'' National Renewable Energy Laboratory (NREL), 2021.

\end{thebibliography}

\appendix

This appendix provides the derivatives of the first-order Taylor expansions necessary for the models \eqref{eq:ac_taylor_opf} and \eqref{eq:dc_taylor_opf}.
\begin{subequations} \label{eq:derivatives}
\begin{align}
& \tfrac{\partial p^f_l}{\partial v_n} = 2G^{ff}_l\tilde{v}_n + \tilde{v}_m(G^{ft}_l\cos\tilde{\theta}_{nm}+B^{ft}_l\sin\tilde{\theta}_{nm}) \\
& \tfrac{\partial p^f_l}{\partial v_m} =  \tilde{v}_n(G^{ft}_l\cos\tilde{\theta}_{nm}+B^{ft}_l\sin\tilde{\theta}_{nm}) \\
& \tfrac{\partial p^f_l}{\partial \theta_{nm}} = \tilde{v}_n\tilde{v}_m(-G^{ft}_l\sin\tilde{\theta}_{nm} +B^{ft}_l\cos\tilde{\theta}_{nm})\\
& \tfrac{\partial q^f_l}{\partial v_n} = -2B^{ff}_l\tilde{v}_n + \tilde{v}_m(G^{ft}_l\sin\tilde{\theta}_{nm}-B^{ft}_l\cos\tilde{\theta}_{nm}) \\
& \tfrac{\partial q^f_l}{\partial v_m} =  \tilde{v}_n(G^{ft}_l\sin\tilde{\theta}_{nm}-B^{ft}_l\cos\tilde{\theta}_{nm}) \\
& \tfrac{\partial q^f_l}{\partial \theta_{nm}} = \tilde{v}_n\tilde{v}_m(-G^{ft}_l\cos\tilde{\theta}_{nm} +B^{ft}_l\sin\tilde{\theta}_{nm})\\
& \tfrac{\partial p^t_l}{\partial v_m} = 2G^{tt}_l\tilde{v}_m + \tilde{v}_n(G^{tf}_l\cos\tilde{\theta}_{mn}+B^{tf}_l\sin\tilde{\theta}_{mn}) \\
& \tfrac{\partial p^t_l}{\partial v_n} =  \tilde{v}_m(G^{tf}_l\cos\tilde{\theta}_{mn}+B^{tf}_l\sin\tilde{\theta}_{mn}) \\
& \tfrac{\partial p^t_l}{\partial \theta_{mn}} = \tilde{v}_m\tilde{v}_n(-G^{tf}_l\sin\tilde{\theta}_{mn} +B^{tf}_l\cos\tilde{\theta}_{mn})\\
& \tfrac{\partial q^t_l}{\partial v_m} = -2B^{tt}_l\tilde{v}_m + \tilde{v}_n(G^{tf}_l\sin\tilde{\theta}_{mn}-B^{tf}_l\cos\tilde{\theta}_{mn}) \\
& \tfrac{\partial q^t_l}{\partial v_n} =  \tilde{v}_m(G^{tf}_l\sin\tilde{\theta}_{mn}-B^{tf}_l\cos\tilde{\theta}_{mn}) \\
& \tfrac{\partial q^t_l}{\partial \theta_{mn}} = \tilde{v}_m\tilde{v}_n(-G^{tf}_l\cos\tilde{\theta}_{mn} +B^{tf}_l\sin\tilde{\theta}_{mn})
\end{align}
\end{subequations}

 





\end{document}